\documentclass[aps,prb,twocolumn,reprint,showpacs,superscriptaddress]{revtex4-1}
\usepackage{amsmath}
\usepackage{amssymb}
\usepackage{graphicx}
\usepackage{hyperref}
\usepackage{url}
\usepackage{color,xcolor}
\usepackage{multirow}
\begin{document}
\title{Topological end states in two-orbital double-exchange model for colossal magnetoresistive manganites}
\author{Yang Li}
\affiliation{Department of Physics, Beijing Normal University, Beijing 100875, China}
\author{Shuai Dong}
\email{Corresponding author: sdong@seu.edu.cn}
\affiliation{Department of Physics, Southeast University, Nanjing 211189, China}
\author{Su-Peng Kou}
\email{Corresponding author: spkou@bnu.edu.cn}
\affiliation{Department of Physics, Beijing Normal University, Beijing 100875, China}
\date{\today}

\begin{abstract}
Manganites are famous mostly for the colossal magnetoresistive effect, which involves the phase separation between ferromagnetic phase and charge-ordered CE-type antiferromagnetic phases. Furthermore, manganites contain some typical magnetic ferroelectrics, e.g. E-type antiferromagnetic $o$-HoMnO$_3$. Here we re-examined these zigzag-winding antiferromagnetic phases (CE-type and E-type antiferromagnets) from the topological perspective. Our theoretical analysis proved that the E-type phase is a weak topological insulator belonging to the $\mathbb{Z}$ class. In momentum space, we classify the symmetries of this phase, and find the three symmetry operators for the chiral, particle-hole, and time-reversal symmetry. The CE-type phase can be described by the Duffin-Kemmer-Petiau algebra, implying that it is a new class of topological insulator and hence extends the existing classification. The corresponding topological end states are demonstrated via numerical calculations, which may implicate the experimental observed ferromagnetic edge states in manganite strips (Nat. Commun. 6, 6179 (2015)) and may play a crucial role in the colossal magnetoresistive effect.
\end{abstract}
\pacs{75.47.Lx, 75.47.Gk, 71.10.Fd}
\maketitle

\section{Introduction}
In recent years, topological matters have become one of the most attractive topics in condensed matter physics.\cite{Hasan:Rmp,Qi:Rmp,Weng:Ap} Protected by the bulk's topological invariants, robust edge (or surface) states can persist against weak perturbations, which may be utilized in power-saving topological electronic devices. Among various topological matters, transition metal oxides with correlated electronic characteristics have drawn more and more attentions not only for its prior properties but for its physical significance beyond the single-electron scenario.\cite{Pesin:Np,Fiete:Sci} For example, recent theoretical studies have predicted Chern insulator or $\mathbb{Z}_2$ topological insulators in various perovskite (111) bilayers,\cite{Xiao:Nc,Yang:Prb,Ruegg:Prb,Wang:Prb15,Weng:Prb} as well as Weyl semi-metal in pyrochlore iridates.\cite{Wan:Prb}

\begin{figure}
\centering
\includegraphics[width=0.48\textwidth]{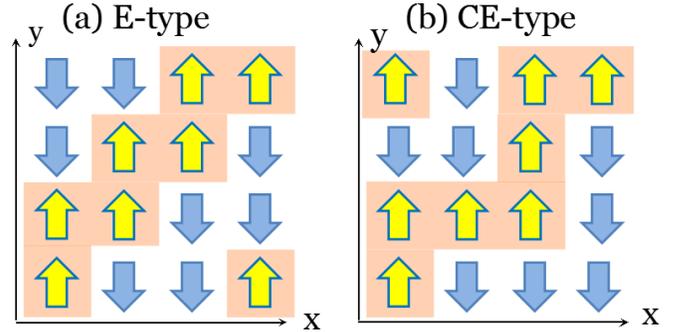}
\caption{(Color online) The spin patterns of $t_{\rm 2g}$ background. (a) The E-type antiferromagnetic phase, which can be stabilized in undoped limit (one $e_{\rm g}$ electron per site). (b) The CE-type antiferromagnetic phase, which can be stabilized around half-doping region (half $e_{\rm g}$ electron per site).}
\label{spin}
\end{figure}

Manganites are typical correlated oxides owning many emergent properties, such as the famous colossal magnetoresistance (CMR). In the past 20 years, enormous theoretical and experimental studies were devoted to understand the physics of manganites.\cite{Tokura:Bok,Dagotto:Bok,Dagotto:Prp,Dagotto:Njp,Tokura:Rpp} It has been widely accepted that the phase competition between ferromagnetic metallic phase and charge ordered phase should be responsible to CMR.\cite{Tokura:Rpp,Sen:Prl,Sen:Prl10} Phase separation, in the scale from nanometers to sub-micrometers, usually exist in various manganites.\cite{Dagotto:Bok,Dagotto:Prp,Dagotto:Njp} The charge ordered phase involved in phase competition and phase separation are mostly the so-called CE-type antiferromagnetic one or its variants.\cite{Tokura:Rpp} The antiferromagnetic pattern of CE phase is constructed by antiferromagnetically-coupled ferromagnetic zigzag chains,\cite{Wollan:Pr} as shown in Fig.~\ref{spin}(b). Other magnetic phases with similar zigzag chains, e.g. the E-type antiferromagnetic one and C$_{1-x}$E$_x$ phase, also exist or have been predicted.\cite{Hotta:Prl04,Dong:Prl} The E-type phase, as shown in Fig.~\ref{spin}(a), was experimental confirmed in undoped narrow-bandwidth manganites, e.g. $o$-HoMnO$_3$,\cite{Munoz:Ic} which is a multiferroic phase.\cite{Sergienko:Prl,Dong:Ap}

In this manuscript, we will prove that both the E-type and CE-type antiferromagnets are also topological insulators. According to the famous ten-fold way classification of topological insulators,\cite{Ryu:Njp,Kitaev:Aipcp} the E-type ferromagnetic chain is a topological insulator belonging to the $\mathbb{Z}$ class. While for the CE-type ferromagnetic chain, we reveal that it is a new kind of topological insulator beyond the present classification of topological insulators, since its effective Hamiltonian can be expressed using a three-dimensional matrix representation of the Duffin-Kemmer-Petiau algebra which is used to describe relativistic spin-$0$ and spin-$1$ particles. A numerical study can simply demonstrate the end (edge) states of both the E-type and CE-type phases. For ideal one-dimensional zigzag chains, the end states are well localized for the E-type phase and quasi-localized for the CE-type one.

Although Hotta \textit{et al.} once proposed the winding number as a real-space topological fingerprint of CE-type and E-type phases to explain their stability,\cite{Hotta:Prl00,Hotta:Rpp} the topology discussed here is somewhat different which is defined in the moment space and associates with the end (edge) state.

\section{Model}
In the following, the standard two-orbital double-exchange model will be studied, which have been repeatedly verified to be a successful model to describe physics of manganites.\cite{Dagotto:Bok,Dagotto:Prp,Hotta:Rpp}. Given the oxygen-octahedral symmetry of the crystalline field, the $3d$-orbitals of Mn split into two groups: the low-lying $t_{\rm 2g}$ triplets and the high-lying $e_{\rm g}$ doublets. Considering the valence of Mn$^{(3+x)+}$ and the high-spin factor due to the strong Hund coupling, the spin-up channel of $t_{\rm 2g}$ orbitals are fully occupied while the spin-up channel of the $e_{\rm g}$ orbitals are partially filled (average $e_{\rm g}$ electron density: $1-x$). The spin-down channels are fully empty for all $3d$ bands. Then the minimal model Hamiltonian without electronic-lattice coupling can be written as:
\begin{equation}
H=-\sum_{i,\mathbf{r},\gamma\gamma'}\Omega_{i,i+\mathbf{r}}t_{\gamma\gamma'}
^{\mathbf{r}}c^\dag_{i,\gamma}c_{i+\mathbf{r},\gamma'}+J_{\rm AF}\sum_{<i,j>}S_{i}\cdot S_{j},
\label{hami}
\end{equation}
where $t_{\gamma\gamma'}^{\mathbf{r}}$ is the nearest-neighbor hopping amplitude between orbital $\gamma$ on site $i$ and the orbital $\gamma'$ on site $i+\mathbf{r}$. In particular, for two dimensional lattices, $t_{aa}^x=t_{aa}^y=3t_{bb}^x=3t_{bb}^y=-\sqrt{3}t_{ab}^x=-\sqrt{3}t_{ba}^x=\sqrt{3}t_{ab}^y=\sqrt{3}t_{ba}^y=3t_0/4$, where $a$ and $b$ are orbitals $x^2-y^2$ and $3z^2-r^2$, respectively. $t_0$ is taken as the energy unit all though this work. $J_{\rm AF}$ is the antiferromagnetic superexchange between the nearest-neighbor $t_{\rm 2g}$ spins ($S_i$ and $S_j$). $\Omega_{i,i+\mathbf{a}}$ is a Berry phase coefficient arising from the spin angles in the infinite Hund coupling limit:\cite{Anderson:Pr,Dagotto:Prp}
\begin{equation}
\Omega_{i,j}=\cos(\frac{\theta_i}{2})\cos(\frac{\theta_j}{2})+\sin(\frac{\theta_i}{2})\sin(\frac{\theta_j}{2})\exp[-i(\phi_i-\phi_j)],
\label{Berry}
\end{equation}
in which ($\theta_i$, $\phi_i$) are the $t_{\rm 2g}$ spin polar and azimuthal angles.

For both the ideal E-type and CE-type antiferromagnetic phases, as sketched in Fig.~\ref{spin}, the $t_{\rm 2g}$ spin backgrounds are constructed by zigzag-like quasi-one-dimensional chains, with ferromagnetic coupling within each chain and antiferromagnetic coupling between nearest neighbor chains. Then one has the maximum hopping ($\Omega_{ij}=1$) along the zigzag chain and the inter-chain hopping is forbidden ($\Omega_{ij}=0$). In other words, theoretically, both the E-type and CE-type phases can be treated as isolated ferromagnetic zigzag chains, as done in previous studies.\cite{Brink:Prl,Hotta:Prl,Hotta:Prl00,Dong:Prb,Dong:Prl}

By comparing the analytic energies calculated using Eq.~\ref{hami} of several candidate phases, a ground state phase diagram (Fig. 7 of Ref.~\onlinecite{Dong:Prb11}) can be obtained straightforwardly, which agrees with the experimental phase diagram very well despite the simplification of model. Both the CE-type and E-type phases emerge in the right areas of parameter space, e.g. undoped and narrow-bandwidth limit for E-type one and half-doped condition for CE-type one. Thus, the current work, based on this successful model, will also lead to practical implication for real materials, instead of pure theoretical meaning of a toy model.

\section{Results: symmetries \& topology}
We see that $t^x_{\mu,\nu}=-t^y_{\mu,\nu}$ for $\mu\neq\nu$. This difference in sign is important. The phase change makes the E-type and CE-type phases topological insulators. To make it clear, we make a unitary transformation:\cite{Hotta:Rpp}
\begin{equation}
\left(
\begin{array}{c}
  \alpha_i \\
  \beta_i
\end{array}
\right)=\frac{1}{\sqrt{2}}
\left(
  \begin{array}{cc}
    1 & i \\
    1 & -i \\
  \end{array}
\right)
\left(
\begin{array}{c}
  c_{i,\gamma} \\
  c_{i,\gamma^{\prime}}
\end{array}
\right).
\label{orbit}
\end{equation}

Then the kinetic energy (the first term in Eq.~\ref{hami}) for an isolated zigzag chain becomes:
\begin{equation}
H_{k}=-\frac{t_0}{2}\sum_{i,\mathbf{r}}(\alpha_i^\dag\alpha_{i+\mathbf{r}}+\beta_i^\dag
\beta_{i+\mathbf{r}}+e^{i\phi_\mathbf{r}}\alpha_i^\dag\beta_{i+\mathbf{r}}+e^{-i\phi_\mathbf{r}
}\beta_i^\dag\alpha_{i+\mathbf{r}}).
\label{kinetic}
\end{equation}
The phase $\phi_\mathbf{r}$ depends on the hopping direction: $-\phi_x=\phi_y=\phi=\pi/3$. In the following, the symmetries and the topology will be discussed based on Eq.~\ref{kinetic}.

\subsection{Symmetries of the E-type chain}
For the E-type chain, the transformed unit cell consists of two nearest neighbor sites, labeled $1$ and $2$. In momentum space, the effective Hamiltonian features a four-band (two orbitals $\times$ two sites) bi-lattice structure. Before solve this four-band Hamiltonian, let's analyse the symmetry first. 

First, the interchange of two sub-lattices in the following way will not change the Hamiltonian (Eq.~\ref{kinetic}):
\begin{equation}
\left(
  \begin{array}{c}
    \alpha_1 \\
    \beta_1 \\
    \alpha_2 \\
    \beta_2 \\
  \end{array}
\right) \qquad \rightarrow \qquad
\left(
  \begin{array}{c}
    \alpha_2 \\
    \beta_2 \\
    \alpha_1 \\
    \beta_1 \\
  \end{array}
\right).
\label{interchange}
\end{equation}
This indicates that the system has a sub-lattice symmetry, representing either a particle-hole exchange or a chirality. Another solution is to make a complex conjugate plus an interchange between orbital $\alpha$ and $\beta$. The operation will not change the Hamiltonian, which indicates that the system overall has a (generalized) time-reversal symmetry.

By choosing a set of proper bases ($\Phi_1^{\pm}=(\alpha_1\pm\beta_1\pm\alpha_2+\beta_2)/2$, $\Phi_2^{\pm}=(\alpha_1\mp\beta_1\pm\alpha_2-\beta_2)/2$), the Hamiltonian matrix in the moment space can be written in a block diagonalized form, as Hotta did:\cite{Hotta:Rpp}
\begin{equation}
h(\mathbf{k})=-\frac{t_0}{2}
\left(
  \begin{array}{cc}
    h^{+}(\mathbf{k}) & 0 \\
    0 & h^{-}(\mathbf{k}) \\
  \end{array}
\right),
\label{diag-ham}
\end{equation}
in which
\begin{eqnarray}
h^+(\mathbf{k})&=&\cos\mathbf{k} \cdot I+2\cos\mathbf{k} \cdot \sigma_z+\sqrt{3}\sin\mathbf{k}\cdot \sigma_x, \nonumber\\
h^-(\mathbf{k})&=&-\cos\mathbf{k} \cdot I+2\cos\mathbf{k} \cdot \sigma_z-\sqrt{3}\sin\mathbf{k}\cdot \sigma_x.
\label{hpm}
\end{eqnarray}
The momentum $\mathbf{k}$ is defined along the zigzag direction.

The chiral symmetry operator can be expressed as:
\begin{equation}
S=\sigma_x\otimes\sigma_x=\left(
    \begin{array}{cc}
      0 & \sigma_x \\
      \sigma_x & 0 \\
    \end{array}
  \right),
\label{chiral}
\end{equation}
which satisfies $SS^\dag=1$ and $S^2=1$. It is straightforward to verify that the Hamiltonian (Eq.~\ref{diag-ham}) has the chiral symmetry $Sh(\mathbf{k})S^{-1}=-h(\mathbf{k})$. This symmetry means that if there is an eigenstate $\psi(\mathbf{k})$ with energy $\varepsilon(\mathbf{k})$, there must be another eigenstate $S\psi(\mathbf{k})$ with energy $-\varepsilon(\mathbf{k})$. Hence, the energy spectrum is symmetrical with regard to the line of energy zero.

Similarly, the particle-hole symmetry operator can be defined as:
\begin{equation}
P=\sigma_x\otimes\sigma_y=\left(
  \begin{array}{cc}
    0 & \sigma_y \\
    \sigma_y & 0 \\
  \end{array}
\right),
\label{par-hol}
\end{equation}
which satisfies $PP^\dag=1$ and $P^T=-P$. Then the particle-hole symmetry of the Hamiltonian can be proved: $Ph(\mathbf{k})P^{-1}=-h(-\mathbf{k})$. Such a particle-hole symmetry means that the spectrum is symmetrical to the zero energy $\Gamma$ point ($\mathbf{k}=0$).

If a system have both the particle-hole and chiral symmetries, then it must have the (generalized) time-reversal symmetry. Here the (generalized) time-reversal symmetry operator is defined as $T=I\otimes \sigma_zK$,
\begin{equation}
T=\left(
    \begin{array}{cc}
      \sigma_z & 0 \\
      0 & \sigma_z \\
    \end{array}
  \right)K,
\label{time}
\end{equation}
which satisfies $TT^\dag=1$ and $T^T=1$. $K$ is complex conjugate. The (generalized) time-reversal symmetry can be confirmed as: $Th(\mathbf{k})T^{-1}=h(-\mathbf{k})$.

According to the famous ten-fold way classification of topological insulators,\cite{Ryu:Njp,Kitaev:Aipcp} the E-type antiferromagnet consisted by one-dimensional zigzag chains is a $\mathbb{Z}$ topological insulator belonging to the BDI class.

In the $\sigma_x-\sigma_z$ plane, direction vectors $\vec{\mathbf{n}}^{\pm}$ can be defined according to Eq.~\ref{hpm}: $\vec{\mathbf{n}}^{\pm}=(\pm\sqrt{3}\sin\mathbf{k},2\cos\mathbf{k})$ for $h^{\pm}(\mathbf{k})$, respectively. These direction vectors rotate in the momentum space. For $h^+(\mathbf{k})$, the orientation of vector $\vec{\mathbf{n}}^+$ rotates clockwise through $2\pi$; for $h^-(\mathbf{k})$, $\vec{\mathbf{n}}^-$ rotates counterclockwise through $2\pi$.

\subsection{Topology of CE-type chain}
Similar procedure can be applied to treat the CE-type chain. After some simple transformations based on Hotta's review,\cite{Hotta:Rpp} the Hamiltonian in momentum space can be written as:
\begin{equation}
h(\mathbf{k})=-t_0
\left(
  \begin{array}{ccc}
    0_{2\times2} & 0 & 0 \\
    0 & M_1 & 0 \\
    0 & 0 & M_2 \\
  \end{array}
\right),
\end{equation}
where the momentum $\mathbf{k}$ is also defined along the zigzag direction. The sub-matrices are:
\begin{eqnarray}
M_1&=&\left(
      \begin{array}{ccc}
        0 & \sqrt{3}\cos\mathbf{k} & 0 \\
        \sqrt{3}\cos\mathbf{k} & 0 & \sin\mathbf{k} \\
        0 & \sin\mathbf{k} & 0 \\
      \end{array}
    \right),    \nonumber \\
M_2&=&
\left(
  \begin{array}{ccc}
    0 & -i\sqrt{3}\sin\mathbf{k} & 0 \\
    i\sqrt{3}\sin\mathbf{k} & 0 & -i\cos\mathbf{k} \\
    0 & i\cos\mathbf{k} & 0 \\
  \end{array}
\right).
\end{eqnarray}
Thus the Hamiltonian decouples into two independent $3\times3$ blocks ($M_1$ and $M_2$) plus the $2\times2$ zero matrix ($0_{2\times2}$).

A more compact expression of $M_1$ and $M_2$ can be written as:
\begin{eqnarray}
M_1&=&\sqrt{3}\cos\mathbf{k}\beta_1+\sin\mathbf{k}\beta_2,  \nonumber  \\
M_2&=&\sqrt{3}\sin\mathbf{k}\beta_2'-\cos\mathbf{k}\beta_1',
\label{Eq16}
\end{eqnarray}
with
\begin{eqnarray}
\beta_1&=&
\left(
  \begin{array}{ccc}
    0 & 1 & 0 \\
    1 & 0 & 0 \\
    0 & 0 & 0 \\
  \end{array}
\right),
\qquad
\beta_2=
\left(
  \begin{array}{ccc}
    0 & 0 & 0 \\
    0 & 0 & 1 \\
    0 & 1 & 0 \\
  \end{array}
\right),   \nonumber \\
\beta_1'&=&
\left(
  \begin{array}{ccc}
    0 & 0 & 0 \\
    0 & 0 & i \\
    0 & -i & 0 \\
  \end{array}
\right)
\quad
\beta_2'=
\left(
  \begin{array}{ccc}
    0 & -i & 0 \\
    i & 0 & 0 \\
    0 & 0 & 0 \\
  \end{array}
\right).
\end{eqnarray}

By defining two more $\beta$'s:
\begin{equation}
\beta_0=\beta_0'=
\left(
  \begin{array}{ccc}
    0 & 0 & -1 \\
    0 & 0 & 0 \\
    1 & 0 & 0 \\
  \end{array}
\right),
\end{equation}
then it is clear that ${\beta_0,\beta_1,\beta_2}$ and ${\beta_0',\beta_1',\beta_2'}$ form three-dimensional matrix representations of the Duffin-Kemmer-Petiau algebra since:
\begin{eqnarray}
\nonumber \beta^\mu\beta^\nu\beta^\sigma+\beta^\sigma\beta^\nu\beta^\mu&=&\beta^\mu
\eta^{\nu\sigma}+\beta^{\sigma}\eta^{\nu\mu},\\
(\beta^{\mu})^3&=&\eta^{\mu\mu}\beta^{\mu}
\end{eqnarray}
where $\eta^{\mu\nu}$ is the Minkowski metric such that $Diag[\eta^{\mu\nu}]=(-1,1,1)$; there is no summation rule on repeated indices. This algebra is a generalized Clifford algebra, which is associated with the Duffin-Kemmer-Petiau theory describing relativistic spin-$0$ and spin-$1$ particles.

As done in above E-type case, the coefficients of Pauli matrices can define a vector direction in the virtual space, then the topology of Hamiltonian can be intuitively judged by observing the windings of this vector direction with changing momentum $\mathbf{k}$. In particular, for the CE-type zigzag chain, the employed algebra is extended from the Clifford algebra to the Duffin-Kemmer-Petiau algebra. For $h_1(\mathbf{k})$ and $h_2(\mathbf{k})$,two vector directions are defined as:
$\vec{\mathbf{n}}_{1}=(\sqrt{3}\cos(\mathbf{k}),\sin(\mathbf{k}))$ in the $\beta_1-\beta_2$ plane and $\vec{\mathbf{n}}_{2}=(-\cos(\mathbf{k}),\sqrt{3}\sin(\mathbf{k}))$ in the $\beta_1'-\beta_2'$ plane. By continuously changing momentum $\mathbf{k}$ through the Brillouin zone, the vector $\vec{\mathbf{n}}_1$ rotates counterclockwise through $2\pi$, and the vector $\vec{\mathbf{n}}_2$ rotates clockwise through $2\pi$.

Besides, Eq.~\ref{Eq16} proves that the topology of CE-type zigzag chain is actually decided by three effective orbital. We can map the CE-type zigzag chain to a topological equivalent chain, by making sure that they have the same form in momentum space. For instance, in momentum space, the following Hamiltonian has the same form as $M_1$:
\begin{equation}
H_{M_1}=-t_0\sum_{i}[\sqrt{3}(a_{i}^{\dag}b_{i+1}+b_{i}^{\dag}a_{i+1})-i(b^{\dag}
_{i}c_{i+1}+c_{i}^{\dag}b_{i+1})+h.c.],
\label{m1}
\end{equation}
in which $a,b,c$ are orbital index. The wave function of electron get a minus sign after hopping along such a loop: $b_{i-1}\rightarrow c_i \rightarrow b_{i+1} \rightarrow a_i \rightarrow b_{i-1}$, which make the system topological nontrivial. The Hamiltonian (Eq.~\ref{m1}) supports the physical argument of the topology of CE type chain.

\section{The topological end states}
A significant feature of a topological insulator is the appearance of an edge state (or an end state for one dimensional cases). In above Sec. III, it has been revealed that the zigzag chains of E-type and CE-type antiferromagnets are both topological insulators. Here the end state will be calculated to further confirm the topology.

\subsection{End states of E-type antiferromagnet}

\begin{figure}
\centering
\includegraphics[width=0.48\textwidth]{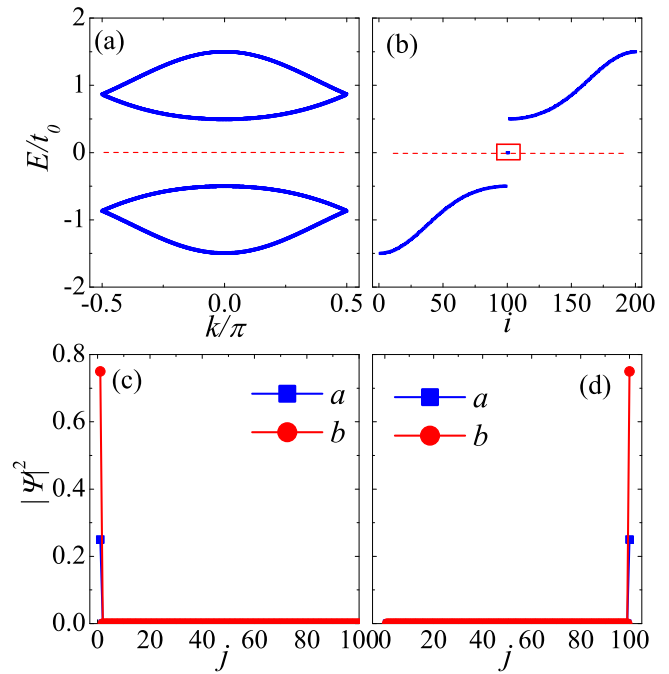}
\caption{(Color online) Electronic structure for E-type antiferromagnetic chain. (a) Energy bands without boundary. (b) Energy spectrum with open boundary conditions. Here the chain length is $100$. $i$: the eigen-energy index. The zero energy point (in red rectangle) corresponds to the end states. Red dot lines in (a-b) denote the Fermi levels for undoped manganites. (c-d) Orbital-resolved electron density distribution for the two end states. Orbital $a$: $d_{x^2-y^2}$; orbital $b$: $d_{3z^2-r^2}$; $j$: the site index.}
\label{Eband}
\end{figure}

Figure~\ref{Eband}(a) shows the band structures of E-type zigzag chain without boundary (i.e. with periodic boundary conditions). To simulate the end state, chains containing even and odd numbers (e.g. $100$ and $101$) of sites with open boundary conditions are adopted. The numerical calculation finds no physical difference regarding the end states between even and odd numbers of sites. As shown in Fig.~\ref{Eband}(b), the mid-gap states at energy zero (circled in red) are double degenerated end states, which do not appear in the calculation with periodic boundary conditions. The edge fact of these two states can be further visualized by calculating their distributions (Fig.~\ref{Eband}(c-d)), which are exactly localized at the two ends of chain. These end states are orbital-related, with different orbital weights. As topological protected end states, their properties (including the energy and distribution) is independent on the length of chains, as confirmed in our numerical calculation (not shown here).

For the ideal E-type AFM without boundary, the chemical potential locates between $t_0$ and $-t_0$ (corresponding to undoped manganites), rendering an insulator. Upon the open boundary, the chemical potential (at zero temperature) is fixed to be $0$, just at the energy level of double-degenerate end states. Thus, the end state will be active to affect the physical properties.

\subsection{End states of CE-type antiferromagnet}

\begin{figure}
\centering
\includegraphics[width=0.5\textwidth]{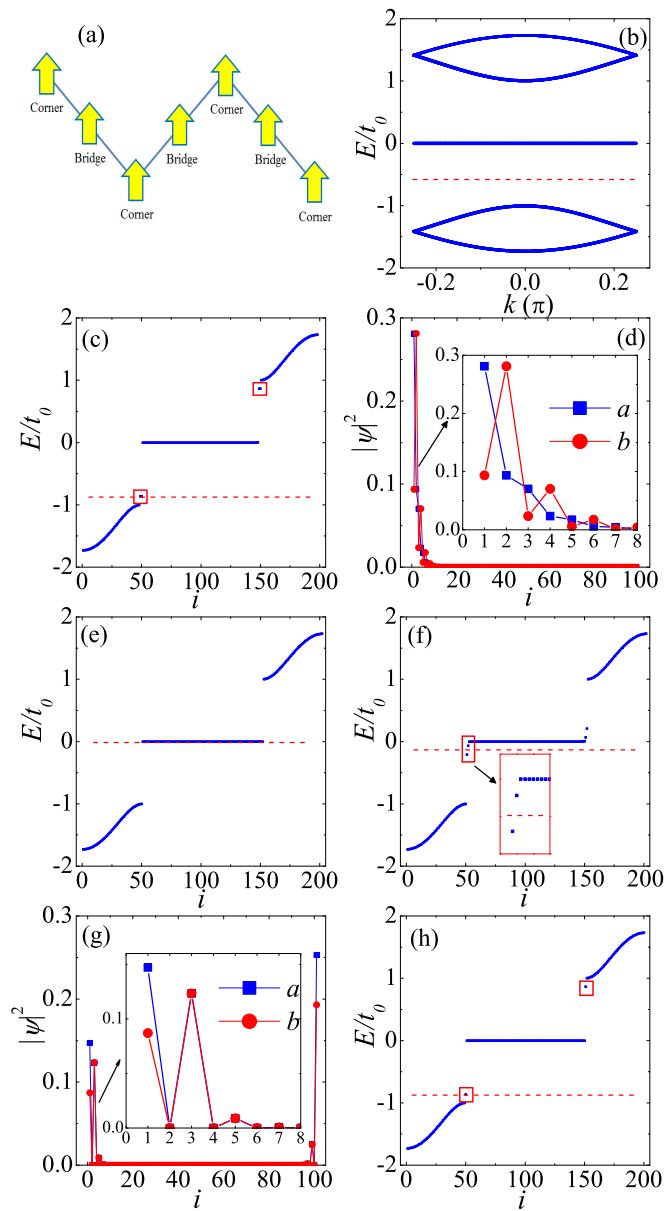}
\caption{(Color online) Electronic structure of the CE-type chain. (a) The corner and bridge sites. (b) Energy bands without boundary. (c-h) Results with open boundary conditions. The chain length is $100+x$. (c-d) $x=1$ with two bridge ends. (c) The energy spectrum. The isolated levels (in red rectangles) are double-degenerate and correspond to the end states. (d) The orbital-resolved electron density distribution for one end state, while another end state shows symmetric distribution at another end (not shown). Insert: a magnified view near the end. (e-g) $x=1$ with two corner ends. (e) Energy bands with open boundary condition. (f) To distinguish the bulk energy states and end states, an imaginary (orbital-dependent) tiny hopping term (here $t'_{aa}=t'_{bb}=0.2t_0,t'_{ab}=t'_{ba}=0.1t_0$ for a better visual effect) is added between two ends. The isolated levels (in red rectangle) correspond to the end states. Insert: a magnified view. (g) The orbital-resolved electron distribution for one of the end states with an imaginary tiny hopping term (here $t'_{aa}=t'_{bb}=0.0002t_0$, $t'_{ab}=t'_{ba}=0.0001t_0$ to get a more precise end state). Insert: a magnified view near the end. (h) The energy spectrum for $x=0$ (one bridge end plus one corner end). Each isolated level (in red rectangles) is singlet. Red dot lines in (b), (c), (e), (f) and (h) denote the Fermi levels for half-doped manganites.}
\label{CEband}
\end{figure}

The end states of CE-type antiferromagnet are more interesting. There are two kinds of sites in the CE-type zigzag chain: the corner one and bridge one,\cite{Dong:Prb} as shown in Fig.~\ref{CEband}(a). The original band structure without boundary is shown in Fig.~\ref{CEband}(b). The energy region can be partitioned into different sections: A: below $-t_0$, B: at zero; C: above $t_0$; D: from $0$ to $-t_0$; E: from $0$ to $t_0$. For the bulk situation (i.e. with periodic boundary conditions), states appear only in the A, B, and C regions, while the D and E are empty (i.e. forbidden bands).

Depending on the odevity of chain length, there are several kinds of situations with open boundary conditions. Since a CE unit consists of four sites, the length of open boundary chain, which can be expressed as $4N+x$ where both $N$ and $x$ ($<4$) are integers, can be commensurate to the period of CE unit (if $x=0$), or not (if $x>0$). In the following, we will analysis these situations one by one.

First, for any odd $x$ with two bridge ends, the end states are clearly seen in energy spectrum. As shown in Fig.~\ref{CEband}(c), there are two isolated energy levels ($\sim\pm0.8t_0$, like shallow impurity levels) which do not show up in the periodic boundary conditions (Fig.~\ref{CEband}(b)). These energy levels are double-degenerated, giving four states in total. These states are indeed end states, confirmed by their spatial distribution (Fig.~\ref{CEband}(d)). Different from the fully localized end states in the E-type antiferromagnet, here the end states are quasi-localized.

Second, for any odd $x$ with two corner ends, the energy spectrum seems to be trivial, without visible ``impurity levels" as the end states. However, by analysing the structure of spectrum (see Table I for more details), it is interesting to find there are overmuch zero energy states. More precisely, there are four more zero energy states comparing with the above two-bridge-end case. Thus, it is possible that the end states hide in the zero energy levels by accident or due to intrinsic symmetry. Even though, these end states can be distinguished from zero energy bulk levels by imposing an additional hopping term between two ends. With increasing amplitude of this hopping term, four branches migrate out from the zero energy position, while the rest $4N$ zero energy levels are unaffected. The wave function distribution of these end states are indeed quasi-localized.

\begin{table}
\caption{Energy spectrum of CE-type zigzag chain with open boundary conditions. The length of chain is $4N+x$ where $N$ and $x$ are integers. The energy region is partitioned into five sections. A: below $-t_0$, B: at zero; C: above $t_0$; D: from $0$ to $-t_0$; E: from $0$ to $t_0$. There is symmetric correspondence between A and C, D and E.}
\centering
\begin{tabular*}{0.48\textwidth}{@{\extracolsep{\fill}}llllll}
\hline \hline
$x$ & Bridge & Corner & A or C & B & D or E\\
\hline
$0$ & $1$ & $1$ & $2N-1$ & $4N$ & $1$ \\
\hline
\multirow{2}{*}{$1$} & $2$ & $0$ & $2N-1$ & $4N$ & $2$\\
\cline{2-6}
& $0$ & $2$ & $2N$ & $4N+2$ & $0$\\
\hline
$2$ & $1$ & $1$ & $2N$ & $4N+2$ & $1$ \\
\hline
\multirow{3}{*}{$3$} & $2$ & $0$ & $2N$ & $4N+2$ & $2$\\
\cline{2-6}
& $0$ & $2$ & $2N+1$ & $4N+4$ & $0$\\
\hline \hline
\end{tabular*}
\label{table2}
\end{table}

Third, for any even $x$, there must be one bridge end plus one corner end. As expected, there are two visible ``impurity levels" from the bridge end while two hidden zero energy levels from the corner end. There are no qualitative physical difference between the $x=0$ (commensurate chain length) and $x=2$ (incommensurate chain length). The distribution of the end states are just simple superposition of two bridge case and two corner case.

In summary, one corner end will induce two symmetrical end state with zero energy, which are related to the nontrivial zero energy band. One bridge end will induce two symmetrical shallow impurity levels, which are related to the nontrivial bulk energy band. Thus, totally, there are four end states in one CE chain with open boundary conditions. All these end states are quasi-localized around the end.

For the ideal CE-type zigzag chain (of $4N$ length), the Fermi level is located between $0$ and $-t_0$ for no-boundary case (corresponding to the half-doped manganites), making the CE phase an insulator too. While for the open boundary conditions, the Fermi level is just between the energy level of lowest energy end states and $0$ for even $x$. While for odd $x$, the expected electron density is a half-integer, and the Fermi level (at zero temperature) is just located at the energy level of lowest energy end states, despite the end type (corner site or bridge site). In this sense, the lowest energy end states are always the topmost valence levels and thus mostly active to determine the physical properties.

\subsection{Robustness of edge states}
All above studies are based on the Hamiltonians in the ideal limit. However, for real manganites there are other important interactions (e.g. electron-lattice couplings, Hubbard-type Coulombian repulsion, next-nearest neighbor hoppings, etc.), which deserve careful analysis to check their effects to end states.

In the following, we will simply check the robustness of edge state upon Jahn-Teller distortion (a kind of electron-lattice couplings), which is very prominent in real manganites. Due to the shortening of lattice axis along the $c$-axis,\cite{Tokura:Rpp} the energy levels of $x^2-y^2$ and $3z^2-r^2$ orbitals will be split, which can be described using the following Hamiltonian:\cite{Dagotto:Prp,Dong:Prb}
\begin{equation}
H_{\rm JT}=\lambda\sum_iQ_{3,i}\tau^z_i.
\end{equation}
For simplify, here only the Jahn-Teller $Q_3$ model is considered, which is uniform for all sites ($Q_{3,i}=Q_3$) (a reasonable assumption). The orbital pseudo-spin operator $\tau^z$ is $c^{\dag}_ac_a-c^{\dag}_bc_b$. Then the coefficient $\lambda Q_3$ determines the energy splitting.

\begin{figure}
\centering
\includegraphics[width=0.5\textwidth]{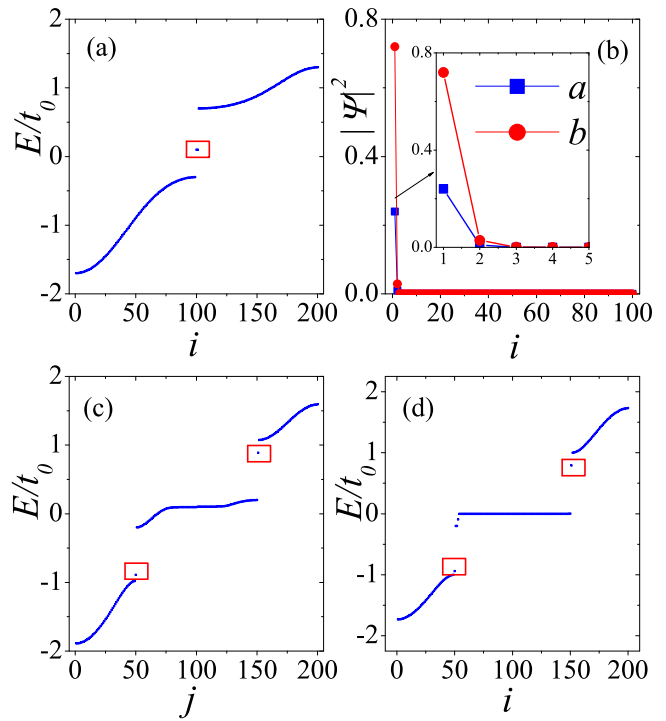}
\caption{(Color online) Electronic structures of E-type and CE-type antiferromagnetic chain against perturbations. The perturbation is in the form of (a-c) uniform Jahn-Teller $Q_3$ mode distortion ($\lambda Q_3=-0.2t_0$) and (d) extra on-site potential ($V_i=-0.2t_0$) applied on two end sites. (a) Energy spectrum for E-type chain with open boundary conditions. (b) Orbital-resolved electron distribution of one end state for E-type chain. (c-d)  Energy spectrums for CE-type chain with open boundary conditions. The length of chain is $100$ with one corner end plus one bridge end. Inserts of (b) and (d) are magnified views near the ends. The red rectangles highlight energy levels of some end states.}
\label{JT}
\end{figure}

For the E-type zigzag chain with such a Jahn-Teller distortion, the typical results are shown in Fig.~\ref{JT}(a-b). Comparing with Fig.~\ref{Eband}(b), the energy spectrum is no more symmetrical to $0$, which means that $H_{\rm JT}$ actually breaks the particle-hole symmetry and time reversal symmetry. Even though, the topology will not be lost and the degeneracy of end states remain near $0$. The electron distribution of end state is shown in Fig.~\ref{JT}(b), which becomes quasi-localized.

Such a Jahn-Teller distortion can also tune the energy spectrum of CE-type chain, as shown in Fig.~\ref{JT}(c). Even though, the end states are still present.

Besides, another perturbation is also tested by applying a chemical potential at the end of chain. \cite{Madsen:Prb} Our numerical calculation confirms the existence of end states for both E-type and CE-type chains (Fig.~\ref{JT}(d)).

In summary of this subsection, the end states show reasonable robustness against some perturbations, such as the Jahn-Teller $Q_3$ distortion and chemical potential at the end of the chain. Of course, more further studies are needed to verify other interactions/perturbations.

\section{Discussion: implication to materials}
In above sections, the symmetry and topology of E-type and CE-type antiferromagnetic chain has been discussed. For real manganites (two-dimensional layers or three-dimensional lattices), the E-type and CE-type magnetic orders are consisted by many independent one dimensional chains. In this case, the edge/surface state of manganites is weak topological state. In real materials, more realistic terms, e.g. lattice distortions, next-nearest-neighbor hoppings, spin-orbit coupling, noncollinear spin canting, and Hubbard-type correlation, will certainly affect the electronic structure. It is interesting to know whether these interactions will modify the topology of CE-type and E-type phases, which needs further careful studies. Considering the complexity of manganites, not only theoretical investigations, but also experimental evidences are needed to clarify the possible topology. Intuitively, the topological end/edge state will probably exist as long as perturbations are small, which was also partially verified in above sections, at least for some kinds of perturbations. Here, it would be helpful to have a brief discussion about the implication of these end/edge states to real materials with CE-type phase. The aim of our discussion is to draw the attention to re-investigate manganites and CMR effect from the topological perspective.

First, a direct relevant experiment is the ferromagnetic edge state observed recently in manganite (La$_{0.325}$Pr$_{0.3}$Ca$_{0.375}$MnO$_3$) strips.\cite{Du:Nc} This manganite is famous for its large scale (up to the micrometer scale) phase separation consisted by charge-ordered phase (mostly believed to be the CE-type or canting CE-type) and ferromagnetic phase.\cite{Uehara:Nat} In Du \textit{et al.}'s experiment, the magnetic force microscopy images clearly indicated that the edges prefer the ferromagnetic tendency.\cite{Du:Nc} A Monte Carlo simulation on two-orbital double-exchange model (done by one of the authors) indeed got the ferromagnetic edge in the nanometer scale which was attributed to the possible seed for experimental observed ferromagnetic edge. Even though, the underlying mechanism remains unclear at that time. Now it is clear that the existence of end states in the CE-phase manganite stripe make the CE-type antiferromagnetic ordering unstable at the edge.

Second, similar situation occurs in small size manganites (e.g. nanoparticle, nanowires, etc.), where ferromagnetic-like surface state usually emerges for charge-ordered (mostly the CE phase or canting CE phase) manganites.\cite{Lu:Apl,Dong:Prb07} A previous model study explained the ferromagnetic tendency for the [001] surface of CE phase,\cite{Dong:Prb08} while it remains an open question for other surfaces. The present study can give a hint to understand these phenomena based on the end/edge state scenario.

Third, previous experiments have found the CE phase is fragile against the B-site substitution.\cite{Markovich:Prb02,Nair:Prl,Banerjee:Prb,Yaicle:Prb,Lu:Prb} Despite many model studies which support the experimental observations,\cite{Pradhan:Prl,Chen:Prb} the real origin is somewhat ambiguous. The robust end states as revealed in our study provide a hint to understand the corrosion of CE phase. Any B-site substitution, no matter magnetic nor nonmagnetic, will make the end states surrounding the ``defect" site, as long as the double-exchange process is blocked there. Considering the quasi-localization of end state in the CE phase, the influenced region is considerable large. In other words, a few B-site substitution can significantly suppress the CE antiferromagnetic order.

Fourth, Brey and Littlewood once proposed a solitonic phase in half-doped manganites, in which each orbital soliton
carried a charge of $\pm e/2$.\cite{Brey:Prl05} In fact, their orbital soliton is closely related to the topological end states of CE phase studied here.

Last but not the least, the end states of CE phase can play an important role in real manganites with phase competitions, which may influence the colossal magnetoresistance. In typical manganites, phase separation with coexisting ferromagnetic/charge-ordering clusters can be modulated by external magnetic fields, giving rise to the colossal magnetoresistance. The edges of CE-type clusters will be active frontier of phase transition and thus play a crucial role in percolative transport. More studies are needed to further clarify the edge states in manganites and their effects in colossal magnetoresistance.

\section{Conclusion}
In the present study, the nontrivial topology of two common antiferromagnetic phases (E-type and CE-type) in manganites have been studied using the standard two-orbital double-exchange model. Our study proved that the E-type phase is a weak topological insulator belonging to the $\mathbb{Z}$ class, while the CE-type phase is a new class of topological insulator characterized by the Duffin-Kemmer-Petiau algebra. The end states associated with the topological bands are also studied, which may be responsible to the new discovered ferromagnetic edge state and some long-standing experimental observations in manganites. The present study not only extends the scope of topological physics in complex oxides, but also provide a new insight to colossal magnetoresistance. Further theoretical and experimental studies are needed to complete the scenario of topological correlated electronics.

\acknowledgments{This work was supported by National Basic Research Program of China (2012CB921704), SRFDP, the Fundamental Research Funds for the Central Universities and National Natural Science Foundation of China (Grant Nos. 11174035 and 51322206).

\bibliographystyle{apsrev4-1}
\bibliography{../../../ref}
\end{document}